\documentclass[a4paper,amsmath,amssymb,aps,pre,reqno,superscriptaddress,twocolumn,floatfix]{revtex4-2}
\usepackage[utf8]{inputenc}
\usepackage[T1]{fontenc}

\usepackage{dcolumn}
\usepackage{mathtools}
\usepackage{bm}
\usepackage{mathrsfs}
\usepackage{subfigure}
\usepackage[document]{ragged2e}

\usepackage{graphicx}
\usepackage[usenames,dvipsnames,table]{xcolor}
\usepackage{comment}
\usepackage{times}

\usepackage[colorlinks=true,linkcolor=Blue,citecolor=Blue,urlcolor=Blue]{hyperref}
\usepackage[capitalise]{cleveref}

\newcommand{\ic}{\mathrm{i}}
\newcommand{\R}{\mathbb{R}}
\newcommand{\T}{\mathbb{T}}
\newcommand{\Z}{\mathbb{Z}}
\newcommand{\eps}{\varepsilon}
\newcommand{\Tb}{\mathbf{T}}

\begin{document}

\title{Collective dynamics on higher-order networks}

\author{Federico Battiston}
\affiliation{Department of Network and Data Science, Central European University, Vienna, Austria}

\author{Christian Bick}
\affiliation{Department of Mathematics, Vrije Universiteit Amsterdam, Amsterdam, The Netherlands}
\affiliation{Institute for Advanced Study, Technical University of Munich, Garching, Germany}
\affiliation{Department of Mathematics, University of Exeter, Exeter, UK}
\affiliation{Mathematical Institute, University of Oxford, Oxford, UK}

\author{Maxime Lucas}
\affiliation{Department of Mathematics and Namur Institute for Complex Systems (naXys), Université de Namur, Namur, Belgium}
\affiliation{Mycology Laboratory, Earth and Life Institute, Université Catholique de Louvain, Louvain-la-Neuve, Belgium}

\author{Ana P. Millán}
\affiliation{Institute ``Carlos I'' for Theoretical and Computational Physics, and Electromagnetism and Matter Physics Department, University of Granada, Granada, Spain}

\author{Per Sebastian Skardal}
\affiliation{Department of Mathematics, Trinity College, Hartford, CT, USA}

\author{Yuanzhao Zhang}
\affiliation{Santa Fe Institute, Santa Fe, NM, USA}

\begin{abstract}
Higher-order interactions that nonlinearly couple more than two nodes are important in many networked systems, and their effects on collective dynamics are increasingly being studied. Here we provide an overview of this rapidly growing field, and of the techniques that can be used to describe and analyze them. We focus in particular on new phenomena and challenges that emerge when nonpairwise interactions are considered. We conclude by discussing open questions and promising future directions on the collective dynamics of higher-order networks.
\end{abstract}

\flushbottom
\maketitle

\thispagestyle{empty}

\justify

\section{Introduction}

The emergence of collective dynamics in networks of interacting dynamical units is a ubiquitous phenomenon in nature and society, and a key signature of many complex systems~\cite{strogatz2014nonlinear,boccaletti2002synchronization,pikovsky2003synchronization,Acebron2005,boccaletti2006complex,arenas2008synchronization}.
Synchronization, where the units evolve in unison, is one of the most striking examples:
Since the first observation by Huygens of the ``strange sympathy between two pendulum clocks'' in 1665, mathematicians and physicists have worked intensively to describe how order can spontaneously emerge as collective dynamics in systems of interacting units.
From traditional synchrony and consensus to collective oscillations and chaos, such collective dynamics are relevant in applications ranging from neuroscience to biology to physics~\cite{roy1994experimental,lotrivc2000synchronization,leloup2003detailed,cumin2007generalising,dumas2010inter,motter2013spontaneous,peleg2024new}.

Traditional models of network dynamical systems, including Kuramoto's famous model~\cite{kuramoto1975self}, 
often consider all-to-all connectivity between units.
These typically do not capture the interaction patterns observed in real-world systems~\cite{arenas2008synchronization}. 
In the last three decades, rapid developments in network theory have led to an avalanche of works devoted to unveiling how the structure of the network influences the emerging collective behavior, investigating the effect of heavy-tailed degree distributions~\cite{gomez2007paths}, short system diameters~\cite{hong2002synchronization}, community structure \cite{arenas2006synchronization} and other widespread properties empirically observed in real-world networks.
Importantly, collective dynamics of coupled units can often be traced to certain network properties, such as abrupt synchronization transitions triggered by degree-frequency correlation~\cite{gomez2011explosive} and cluster synchronization induced by network symmetries~\cite{stewart2003symmetry,pecora2014cluster,Aguiar2018,Snchez-Garca2020}.

Despite these advances, traditional networks cannot capture nonpairwise (also known as higher-order or polyadic) ties, where more than two units are jointly interacting~\cite{battiston2020networks,lambiotte2019networks, bianconi2021higher,bick2023higher}. Indeed, such systems are better described by higher-order modeling frameworks, such as simplicial complexes or hypergraphs, where hyperedges encode structured interactions among any number of units.
Recently, a stream of literature has been pointing out the importance of higher-order interactions in both natural and man-made systems, showing that they can drastically reshape the collective dynamics of a system, as extensively reviewed in Refs.~\cite{bick2023higher,battiston2021physics,boccaletti2023structure,majhi2022dynamics}. 
Here start with synchronization and oscillatory dynamics as the unifying scaffold to review new phenomena that emerge in higher-order networks, since they are some of the most actively researched dynamical processes on higher-order networks and often encompass other processes such as consensus~\cite{neuhauser2020multibody,neuhauser2021consensus,hickok2022bounded,kim2025competition}, diffusion~\cite{millan2021local}, and random walk~\cite{carletti2020random} as special cases. 
Where appropriate, we will frame key results in the context of general dynamical processes.
For in-depth discussions on specific processes, such as contagion and topological signals, we refer the readers to other recent reviews~\cite{ferraz2024contagion, millan2025topology}. 

\begin{figure*}
    \centering
    \includegraphics[width=0.99\linewidth]{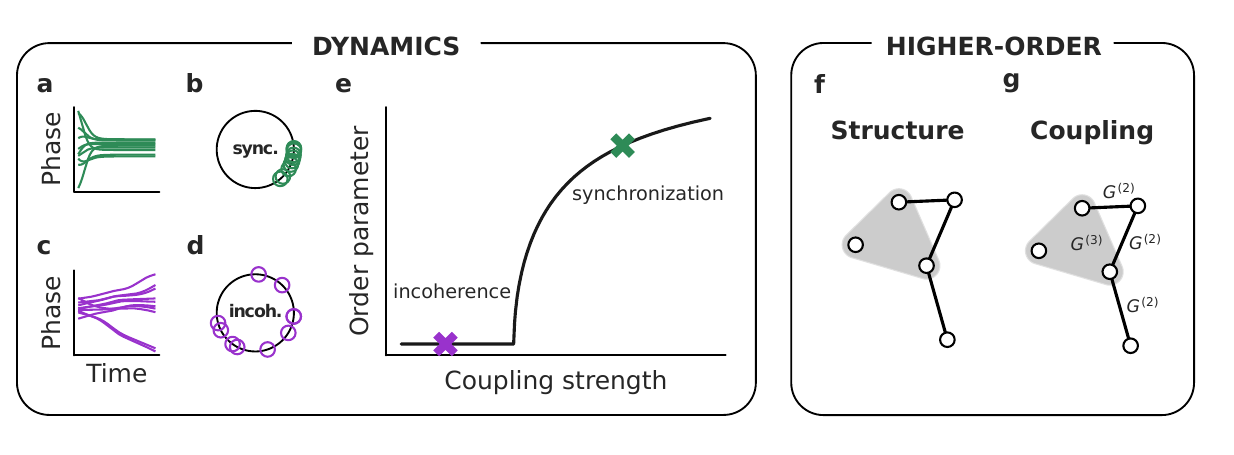}
    \caption{Dynamics and higher-order networks. Coupled dynamical units such as Kuramoto oscillators can exhibit ordered and disordered states such as (a-b) synchronization, and (c-d) incoherence. (e) Typically, in networks there is a continuous transition from incoherence to synchronization as the coupling strength is increased. Adding higher-order (f) structure, and (g) coupling to the network can fundamentally change the dynamics, which is the focus of this review.}
    \label{fig:intro}
\end{figure*}

An overview of the content of the review is summarized in \cref{fig:intro}. 
First, we focus on how higher-order structures can influence collective dynamics.
We start by describing synchronization phenomena in generalized Kuramoto oscillators with polyadic ties, how to provide analytical descriptions of such systems, and the emergence of new phenomena such as explosive transition, multistability, and collective dynamics beyond traditional synchronization.
Second, we go beyond Kuramoto dynamics and discuss more general node dynamics with nonpairwise interactions.
For oscillatory intrinsic dynamics, we connect general oscillator dynamics back to generalized Kuramoto oscillators through phase reduction theory:
Higher-order phase interactions may capture indirect interactions between limit-cycle oscillators.
Third, we go in the opposite direction and discuss how dynamics can inform us about the higher-order organization of a system.
In particular, we discuss how hypergraphs and simplicial complexes can be reconstructed from time-series data, and show that observations of dynamics unfolding can help reduce the model complexity on higher-order networks.
Finally, we go beyond nodal dynamics and characterize collective phenomena in dynamical systems where state variables are not only associated with nodes, but also edges and hyperedges.
We conclude by presenting open questions and promising directions for future research in the fast-developing and vibrant field of dynamical higher-order networks.

\section{Kuramoto Oscillators with Higher-Order Interactions}
\label{sec:PhaseSync}

The Kuramoto model~\cite{kuramoto1975self,kuramoto1984chemical,strogatz2000kuramoto} describes the evolution of $n$~oscillators with states $\theta_i\in\T=\R/2\pi\Z$ and intrinsic frequencies $\omega_i\in\R$ for $i\in\{1,\dotsc,n\}$.
To understand the effect of nonpairwise interactions on synchronization dynamics, we start with a generalized Kuramoto model~\cite{tanaka2011multistable,skardal2019abrupt}
\begin{equation}
\begin{split}
    \dot{\theta}_i = \omega_i & + \sigma \sum_{j=1}^{n} A_{ij} \sin(\theta_j-\theta_i) \\
    & + \sigma_\triangle^{(\rm s)} \sum_{j,k=1}^{n} B_{ijk} \sin(\theta_j+\theta_k-2\theta_i) \\
    & + \sigma_\triangle^{(\rm as)} \sum_{j,k=1}^{n} C_{ijk} \sin(2\theta_j-\theta_k-\theta_i), 
\end{split}
\label{eq:kuramoto}
\end{equation}
where in addition to the pairwise coupling---described by the coefficients~$A_{ij}$ of a (weighted) adjacency matrix and coupling strength~$\sigma$---there are nonlinear interactions between triplets of oscillators determined by the (weighted) adjacency tensors~$B_{ijk}$ and~$C_{ijk}$ of coupling strengths~$\sigma_\triangle^{(\rm s)}$ and $\sigma_\triangle^{(\rm as)}$, respectively.
These tensors correspond to two distinct types of coupling functions~\cite{stankovski2015coupling}:
The triplet interaction $\sin(\theta_j+\theta_k-2\theta_i)$ is \emph{symmetric} in the inputs~$j$ and~$k$ while the \emph{asymmetric} coupling function $\sin(2\theta_j-\theta_k-\theta_i)$ is not.
For global coupling, \cref{eq:kuramoto} is related to phase dynamics with nonlinear mean-field coupling~\cite{rosenblum2007self}.
The coupling through phase differences induces a rotational symmetry where $\alpha\in\T$ acts as a common phase shift to all oscillators, $\theta\mapsto\theta+\alpha$.

On the one hand, generalized Kuramoto equations such as~\eqref{eq:kuramoto} can be considered as models \emph{per se} to understand synchronization dynamics~\cite{matheny2019exotic,skardal2021higher}.
Varying parameters such as $\sigma_\triangle^{(\rm s)}$ and $\sigma_\triangle^{(\rm as)}$ independently allows one to analyze how the nonpairwise interactions shape the synchronization dynamics.
On the other hand, equations of this type can be derived from phase reductions, as we outline in~\cref{sec:phase_reduction}.
The resulting equations link to more general nonlinear oscillators but nonpairwise interactions of
different types typically arise simultaneously and depend on the physical model parameters~\cite{leon2019phase}.
We first consider oscillators with identical intrinsic frequencies as it allows us to isolate the influence of coupling structures.
We then turn to nonidentical oscillators, which display explosive transitions and extensive multistability when coupled through nonpairwise interactions.

\begin{figure*}
    \centering
    \includegraphics[width=0.9\linewidth]{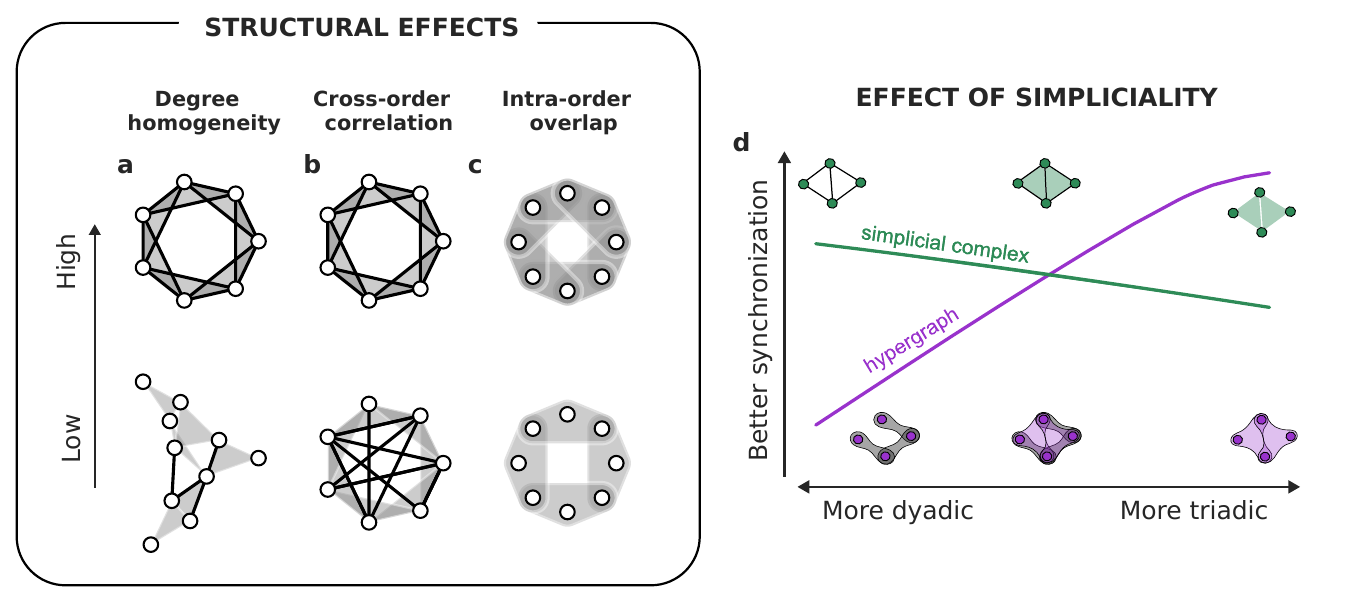}
    \caption{Effect of higher-order structure on dynamics. Various structural properties affect dynamics such as (a) degree homogeneity, (b) cross-order degree correlation, defined as the correlation between degree sequences from different orders, and (c) intra-order overlap, measuring the overlap between hyperedges of the same order. For example, (d) simpliciality (how close is a hypergraph from becoming a simplicial complex), which affects both degree homogeneity and cross-order degree correlation, can change whether higher-order interactions stabilize or destabilize synchronization.}
    \label{fig:identical}
\end{figure*}

\subsection{Identical oscillators and the role of hypergraph structure}
\label{sec:IdPhaseOsc}

A natural starting point is to analyze \emph{phase synchrony} $\theta_i(t)=\theta_j(t)$, which is invariant for the generalized Kuramoto oscillators~\eqref{eq:kuramoto}
if the intrinsic frequencies of the phase oscillators are identical, $\omega_i=\omega$.
Is this state (linearly) stable and, importantly, how does this depend on the network structure given by~$A_{ij}$, $B_{ijk}$, and~$C_{ijk}$?
We will discuss more complex synchronization patterns later in the section and in \cref{sec:sync-patterns}.

In pairwise networks, identical frequencies allow us to perform linear stability analysis using the graph Laplacian $L_{ij} = K_i \delta_{ij}  - A_{ij}$. 
Full phase synchrony is stable if and only if all nontrivial eigenvalues of the graph Laplacian are negative.
In higher-order networks, we can use a natural generalization of the graph Laplacian, the \textit{multiorder Laplacian}~\cite{lucas2020multiorder, gambuzza2021stability}. 
It is defined as
\begin{equation}
  \mathbf{L}^{(2, \mathrm{ mul})} =  
  \sigma \mathbf{L}^{(1)} 
  + \sigma_\triangle^{(\rm s)} \mathbf{L}^{(2, \rm s)}  
  +  \frac12  \sigma_\triangle^{(\rm as)}  \mathbf{L}^{(2, \rm as)} ,
\label{eq:multiorder_laplacian}
\end{equation}
where the generalized Laplacians at each order $d$ are defined as 
$ L^{(d)}_{ij} = d K^{(d)}_i \delta_{ij}  - A^{(d)}_{ij} $, 
in terms of the generalized degrees~$\bm{K}^{(d)}$ and generalized adjacency matrix $\mathbf{A}^{(d)}$ of order~$d$. 
For an arbitrary hypergraph, the generalized degree~$K^{(d)}_i$ of node~$i$ is the number of hyperedges of order $d$ of which it is a part. 
For example, $K^{(2, \rm s)}_i = \frac12 \sum_{j,k=1}^n B_{ijk}$ and similarly in the asymmetric case. 
Similarly, the generalized adjacency matrix $A^{(d)}_{ij}$ between nodes $i$ and $j$ is the number of hyperedges of order $d$ of which both $i$ and $j$ are part, for example $A^{(2, \rm s)}_{ij} = \sum_{k=1}^n B_{ijk}$. 
The multiorder Laplacian satisfies the properties expected from a Laplacian: It is positive semidefinite and its rows (columns) sum to zero. 
The multiorder Laplacian is a powerful tool because it extracts all the information relevant for synchronization stability from the tensors $A_{ij}$, $B_{ijk}$, $C_{ijk}$ and packages them into a single Laplacian matrix.
It also naturally relates to more general approaches to determine stability of (cluster) synchrony~\cite{salova2021cluster,salova2021analyzing,Aguiar2020,zhang2020symmetry}.

A key question is which network characteristics promote synchronization and which ones impede it~\cite{barahona2002synchronization}.
The tensors~$B_{ijk}$ and~$C_{ijk}$ that capture the nonpairwise phase interactions are typically adjacency tensors of \textit{hypergraphs} or \textit{simplicial complexes}, two commonly used representations for higher-order network interactions.
Hypergraphs are the most general representation, and simplicial complexes additionally require \textit{downward closure} to be satisfied: For any $d$-body interaction, all ($d-1$)-body interactions of the same nodes must also be included~\cite{battiston2020networks}.
In most cases, the two representations have been used interchangeably, and the choice for one or the other was often motivated by technical convenience---for example, topological data analysis and Hodge decomposition require simplicial complexes~\cite{giusti2016two,patania2017topological}.

However, higher-order interactions under these two representations can affect dynamics very differently---they enhance synchronization in a wide range of hypergraphs but consistently impede synchronization in simplicial complexes~\cite{zhang2023higher}, as illustrated in \cref{fig:identical}d. 
Using the multiorder Laplacian framework, one can trace the origin of these divergent trends to the dramatically different degree heterogeneities present in the two representations.  
In particular, due to the downward closure condition in simplicial complexes, hyperedges are disproportionately attached to nodes that are already well connected through pairwise edges.
This rich-get-richer effect makes simplicial complexes structurally highly heterogeneous, which is the opposite to what happens in typical hypergraphs.
Beyond synchronization, the difference in structural heterogeneity between hypergraphs and simplicial complexes also has significant consequences in many other dynamical processes, such as complex contagions~\cite{burgio2024triadic}.

Beyond the relatively crude distinction between hypergraphs and simplicial complexes, more granular structural features of higher-order networks can also significantly influence dynamical processes (\cref{fig:identical}a--c).
Some of these properties are naturally extended from networks, such as generalized degree heterogeneity~\cite{landry2020effect,zhang2023higher,lucas2026reducibility}, while others are intrinsically higher-order and have no counterparts in networks, such as cross-order degree correlation~\cite{landry2020effect,zhang2023higher,lucas2026reducibility}, and hyperedge overlap~\cite{lamata2025hyperedge}.
The hyperedge overlap can be further disentangled into inter-order hyperedge overlap (nestedness)~\cite{larock2023encapsulation,kim2023contagion,skardal2023multistability,adhikari2023synchronization,kim2024higher,burgio2024triadic, malizia2025disentangling} and intra-order hyperedge overlap~\cite{malizia2023hyperedge}.
For example, it was found that higher hyperedge overlap leads to earlier but smaller outbreaks in contagion dynamics~\cite{kim2023contagion,kim2024higher,burgio2024triadic,malizia2025disentangling}, can hinder synchronizability~\cite{lamata2025hyperedge}, and that lower cross-order degree correlation can suppress bistability~\cite{landry2020effect}. While nodes belonging to the same community are known to synchronize more easily in pairwise networks~\cite{arenas2006synchronization}, a systematic investigation of the effect of higher-order modular structure~\cite{eriksson2021choosing, contisciani2022inference, ruggeri2023community} on synchronization is still to be undertaken.

Although full synchrony is one of the most studied collective states on higher-order networks, there are other synchrony patterns that can emerge in generalized Kuramoto equations~\eqref{eq:kuramoto}.
These arise naturally if the underlying network has symmetries (irrespective of higher-order interactions):
For identical oscillators, network symmetries for permutation of nodes translate into symmetries of the coupled dynamical system~\cite{Golubitsky2002,Snchez-Garca2020}, which yields invariant states~\cite{Ashwin1992,stewart2003symmetry} such as cluster synchrony patterns.
Thus, finding symmetries can provide an essential tool to identify cluster synchrony patterns~\cite{pecora2014cluster}, even in networks with higher-order interactions~\cite{bick2016chaos}.

As a concrete example, full phase synchrony is the fully symmetric state in all-to-all coupled networks while ring-like networks with cyclic symmetries naturally support twisted states (rotating wave solutions where the oscillator phases wrap around the circle in a linear fashion)~\cite{wiley2006size,zhang2021basins}.
For identical Kuramoto oscillators with pairwise couplings, twisted states (which includes full synchrony) are the only stable patterns in ring-like topologies~\cite{delabays2017size,diaz2024exploring}.
Nonpairwise interactions are one way to break the gradient structure in such systems~\cite{Sclosa2025}, which allows for a larger variety of stable collective dynamics such as rotating waves, anti-phase clusters, chimeras, and disordered states~\cite{komarov2015finitesizeinduced,kundu2022higher,bick2023hopf,zhang2024deeper}.
As a result, higher-order interactions can simultaneously increase multistability (making basins of attraction ``smaller'') as well as linear stability (a ``deeper'' basin of attraction)~\cite{zhang2024deeper}.
For example, a twisted state can become linearly more stable but at the same time becomes impossible to reach from random initial conditions due to its basin shrinking drastically (squeezed by other newly-created attractors).

The global organization of phase space gives insights into collective phenomena beyond stability of specific synchronized states.
On the one hand, nonpairwise phase interactions can facilitate the emergence of chaotic dynamics~\cite{bick2016chaos} or collective dynamics organized by global objects such as heteroclinic cycles~\cite{bick2018heteroclinic,bick2019heteroclinica} and networks~\cite{bick2019heteroclinicb}.
On the other hand, for specific higher-order interactions, the dynamics may be reduced to low-dimensional submanifolds through the Watanabe--Strogatz reduction~\cite{watanabe1993integrability,watanabe1994constants} (for finite~$n$) and the related Ott--Antonsen reduction~\cite{ott2008low,ott2009long,Pikovsky2011,Bick2018c} (in the limit of $n\to\infty$).
Specifically, higher-order interactions through nonlinear mean-field coupling can give rise to quasiperiodic collective dynamics~\cite{Rosenblum2007}.
For finite networks, the Watanabe--Strogatz reduction gives information about clustering among oscillators~\cite{komarov2015finitesizeinduced,gong2019lowdimensional}.
In the limit of infinitely many oscillators, an explicit bifurcation analysis allows one to understand how higher-order interactions stabilize or destabilize twisted states~\cite{bick2023phase,bick2023hopf}.

\subsection{Nonidentical oscillators and explosive transitions}

\begin{figure*}
    \centering
    \includegraphics[width=0.99\linewidth]{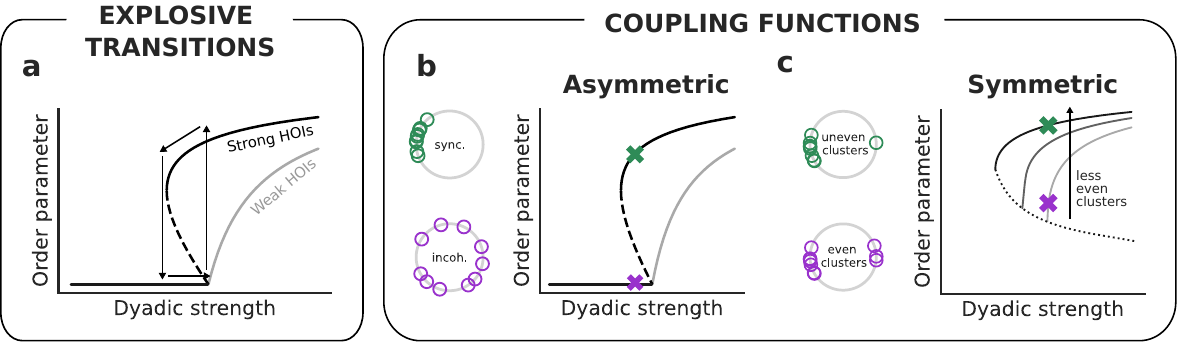}
    \caption{Higher-order interactions can induce explosive transitions and multistability. (a) Hysteresis loop can develop between order parameter and dyadic coupling strength when sufficiently strong higher-order interactions are present. (b--c) The choice of nonpairwise coupling function, for example, asymmetric (\cref{eq:explosive:01}, panel b) or symmetric (\cref{eq:explosive:03}, panel c), can lead to distinct stable states and bifurcation diagrams.}
    \label{fig:transitions}
\end{figure*}

In large ensembles of heterogeneous dynamical systems, a central question is how individual heterogeneity competes with coupling between units to promote collective behavior. In the case of phase oscillators, heterogeneity is captured by non-uniformity in the natural frequencies, which are typically assumed to be drawn from some distribution $\Omega(\omega)$. In the classical Kumamoto model~\cite{strogatz2000kuramoto}, the interplay between the spread in natural frequencies and dyadic sinusoidal coupling yields a phase transition from incoherence to coherence known as the onset of synchronization, where the magnitude $r$ of the complex order parameter $z=re^{\ic\phi}=(1/n)\sum_{j=1}^n e^{\ic\theta_j}$ increases from $r\approx0$ (incoherence) to $r>0$ (partial synchrony).
Incorporating higher-order interactions of different kinds yields rich dynamics that include explosive synchronization transitions and multistability, as also observed in contagion dynamics~\cite{iacopini2019simplicial,ghosh2023dimension}. 

The effects of higher-order interactions can best be understood via a mean-field model.
A generalization of the Kuramoto model with asymmetric triadic coupling as well as the tetradic coupling~\cite{skardal2020higher} is
\begin{equation}\label{eq:explosive:01}
    \begin{split}
\dot{\theta}_i &= \omega_i + \frac{\sigma}{n}\sum_{j=1}^n\sin(\theta_j-\theta_i)\\
&\qquad + \frac{\sigma_\triangle^{(\rm as)}}{n^2}\sum_{j,k=1}^n\sin(2\theta_j-\theta_k-\theta_i)\\
&\qquad + \frac{\sigma_{\Box}}{n^3}\sum_{j,k,l=1}^n\sin(\theta_j+\theta_k-\theta_l-\theta_i),
\end{split}
\end{equation}
where the interaction strengths are appropriately scaled with the system size.
Using the Ott--Antonsen reduction~\cite{ott2008low,ott2009long} in the limit $n\to\infty$ and assuming that natural frequencies are drawn from a Lorentzian distribution with spread~$\varsigma$, one can show that the amplitude~$r=|z|$ of the order parameter evolves according to
\begin{align}
\dot{r}=-\varsigma r+\frac{\sigma}{2}r(1-r^2) + \frac{\sigma_{\triangle}^{(\rm as)}+\sigma_{\Box}}{2}r^3(1-r^2).\label{eq:explosive:02}
\end{align}
Importantly, \cref{eq:explosive:02} reveals that in terms of the macroscopic dynamics, higher-order interactions manifest as purely nonlinear terms and do not affect the linear stability of the incoherent state $r=0$. 
Moreover, this nonlinearity gives rise to a subcriticality for sufficiently large higher-order coupling ($\sigma_{\triangle}^{(\rm as)}+\sigma_{\Box}>2\varsigma$), in which a hysteresis loop forms and a region of bistability between incoherence and synchronization appears, as illustrated in Fig.~\ref{fig:transitions}a. 
Thus, by increasing and decreasing the dyadic coupling strength $\sigma$, the system undergoes explosive transitions between incoherence and partial synchronization, see Fig.~\ref{fig:transitions}b for an illustration of these coexisting states.

If the mean-field coupling in the generalized Kuramoto equations~\eqref{eq:explosive:01} is replaced with nontrivial coupling structures, higher-order interactions of the same kind continue to yield bistability and explosive transitions via a hysteresis loop~\cite{adhikari2023synchronization}. Moreover, the hysteresis loop induced by higher-order interactions can be compounded into multiple tiers by adding time delays between the oscillators~\cite{skardal2022tiered}.  Interestingly, while higher-order interactions do not affect the linear stability of the incoherent state $r=0$ in \cref{eq:explosive:01}, this can change when inertial terms are present~\cite{sabhahit2024prolonged}.
Additionally, incorporating community structure into the coupling leads to added multistability, resulting in anti-phase synchronized and skew-phase synchronized states emerging alongside incoherent and synchronized states~\cite{skardal2023multistability}.
Adding another layer of complexity, a few recent studies have considered mobile oscillators (rather than static), which can produce rich dynamics with spatial patterns and bistability~\cite{anwar2024collective,anwar2025twodimensional,hu2025effect}. 
Finally, recent work has also extended pinning control methods to higher-order networks to promote~\cite{Chen2021controllability,delellis2023pinninglinear,dellarossa2023emergence,rizzello2024pinning,muolo2024pinning,wang2024synchronization,Li2024synchronization,Xia2024pinning} or suppress synchrony~\cite{moriame2024hamiltonian}, to account for these changes in the dynamics.

In the specific case of symmetric triadic coupling, even when analyzed on its own, analytical results are more difficult to come by~\cite{skardal2019abrupt}. 
Consider
\begin{align}
\dot{\theta}_i = \omega_i + \frac{\sigma_\triangle^{(\rm s)}}{n^2}\sum_{j,k=1}^n\sin(\theta_j+\theta_k-2\theta_i),\label{eq:explosive:03}
\end{align}
where the $\pi$-periodicity (instead of $2\pi$-periodicity) of the coupling function in the direction of~$\theta_i$ hinders our ability to derive a closed amplitude equation for the macroscopic dynamics using the Ott--Antonsen ansatz. 
However, some progress can be made by, similar to Ref.~\cite{skardal2011cluster}, applying the Ott--Antonsen ansatz to the symmetric portion of the distribution of oscillators, resulting in an amplitude equation for the amplitude~$r_2$ of the generalized order parameter $z_2=(1/n)\sum_{j=1}^n e^{2\ic\theta_j}$. 
This partial dimensionality reduction still requires a self-consistency analysis but reveals the emergence of cluster states where oscillators become entrained to one of two subpopulations at opposite angles. 
More specifically, the asymmetry between clusters yields extensive multistability in possible configurations~\cite{tanaka2011multistable}, with larger asymmetry (i.e., stronger unevenness between clusters) giving rise to a larger value of~$r$~\cite{xu2020bifurcation,xu2021spectrum}, as illustrated in Fig.~\ref{fig:transitions}c. 
Interpreting oscillators in each cluster as a~$0$ or~$1$ leads to the ability of such systems to have memory and store complicated strings of information~\cite{skardal2020memory}.

\section{General network dynamics with higher-order interactions}

How do higher-order interactions shape the emergent dynamics of coupled dynamical nodes whose states are not simply given by a one-dimensional phase variable (e.g., Kuramoto oscillators) but live in a more general state space?
In this section, we consider a general class of coupled dynamical nodes~\cite{carletti2020dynamical,Aguiar2020}, where the state  $x_i\in\R^{d_i}$ of node~$i\in\{1, \dotsc, n\}$ evolves according to
\begin{equation}
\begin{split}
    \dot{x}_i = F_i(x_i) & +\sigma \sum_{j=1}^{n} A_{ij} G^{(2)}(x_j;x_i) \\
    & +\sigma_{\triangle} \sum_{j,k=1}^{n} B_{ijk} G^{(3)}(x_j,x_k; x_i) 
    + \dotsb
\end{split}
\label{eq:coupled-oscillators-HOI}
\end{equation}
Here~$F_i$ determines the intrinsic dynamics of node~$i$, the tensors~$A_{ij}, B_{ijk},\dotsc$ the coupling structure, and~$G^{(q)}$ the functional form of the interactions of order~$q$ between nodes (which are generally assumed to be nonlinear).
The network dynamics~\eqref{eq:coupled-oscillators-HOI} encompass model equations for a wide range of physical systems including consensus dynamics~\cite{neuhauser2020multibody,neuhauser2021consensus} or mean-field approximations of contagion processes~\cite{iacopini2019simplicial,de2019social,landry2020effect,st2021master}.
We focus here on general collective dynamical phenomena; see~\cite{ferraz2024contagion} for an overview of recent results specific to contagion dynamics.

Below, we first discuss different collective dynamics that can arise from general intrinsic dynamics~$F_i$, how to find the best coordinates to analyze their stability, and the challenges introduced by higher-order interactions.
To connect to the results of the previous section, we then assume that~$F_i$ give rise to stable limit cycles and discuss how generalized Kuramoto equations such as~\eqref{eq:kuramoto} relate to the general system~\cref{eq:coupled-oscillators-HOI} through phase reduction.

\subsection{Dimensionality reduction for general node dynamics}

\label{sec:sync-patterns}

For identical node dynamics~$F_i$, the master stability function~(MSF) formalism~\cite{pecora1998master} is the prevailing paradigm to asses linear stability of synchrony as a basic example of collective dynamics of nodes with general dynamics.
It offers crucial insights connecting network structure and collective dynamics.
The basic idea of MSF is to decouple local dynamics from network structure.
In particular, by switching to coordinates that diagonalize the coupling matrix representing the network, one can keep the stability problem low-dimensional for arbitrarily large networks.
The effect of the network structure is then encoded through a set of eigenvalues, which is independent from the node dynamics and coupling functions.

For systems with nonpairwise interactions, similar approaches in the spirit of MSF have been proposed \cite{Mulas2020,gambuzza2021stability}.
However, there are a few challenges in fully adapting MSF from networks to hypergraphs.
First, unlike the adjacency matrix or Laplacian matrix for networks, the coupling structure in \cref{eq:coupled-oscillators-HOI} is described by tensors.
Fortunately, Ref.~\cite{gambuzza2021stability} showed that, in the context of linear stability analysis, one does not have to deal with tensors directly.
Specifically, the tensor used to describe each order of interactions can be reduced to a corresponding Laplacian matrix, which has been referred to as generalized Laplacians (same definition as the ones in \cref{eq:multiorder_laplacian}). 
Unlike the case of Kuramoto oscillators, however, these generalized Laplacians cannot be combined to form a single multiorder Laplacian.
This is because the synchronization state is no longer a fixed point---the chaotic synchronization dynamics require different coupling functions to be treated separately.

The presence of more than one Laplacian matrix in the stability problem introduces another unique challenge.
There is a notion of ``optimal networks'' for synchronization in networks, which does not depend on the details of the node dynamics or interaction function~\cite{nishikawa2003heterogeneity,nishikawa2006maximum,nishikawa2006synchronization}.
These optimal networks maximize synchronization stability and are characterized by fully degenerate Laplacian eigenvalues (excluding the trivial zero eigenvalue corresponding to perturbing all oscillators in exactly the same way)~\cite{nishikawa2010network}. 
Can we perform the same structural optimization for hypergraphs in \cref{eq:coupled-oscillators-HOI}? Is there an equivalent notion of ``optimal hypergraphs'' for synchronization?
These are outstanding questions because, unlike the master stability functions for networks, full eigendecomposition is usually not possible for hypergraphs due to the generalized Laplacians not commuting with each other, which significantly increases the dimensionality of the stability problem.

Generalizing \cref{eq:coupled-oscillators-HOI} even further, the MSF framework has also been extended to hypergraphs or simplicial complexes with additional temporal~\cite{anwar2024global}, multilayer~\cite{rajwani2023tiered,pal2024global}, adaptive~\cite{rathore2023synchronization,anwar2024synchronization}, or non-reciprocal~\cite{gallo2022synchronization} structures.
As we move up in terms of model complexity, however, it remains an outstanding challenge how to effectively reduce the dimensionality of the system for tractable analyses that can offer new insights.

Beyond full synchronization, cluster synchronization states can also be observed in \cref{eq:coupled-oscillators-HOI}, where the system spontaneously breaks into multiple clusters of oscillators that are only synchronized internally \cite{stewart2003symmetry,aguiar2011dynamics}.
As in \cref{sec:IdPhaseOsc}, structural features of the network give flow invariance of spaces in~\cref{eq:coupled-oscillators-HOI} that correspond to cluster synchrony patterns.
This includes classical symmetries~\cite{Snchez-Garca2020,golubitsky2005patterns,pecora2014cluster} or generalized symmetries (such as quiver- or fibration symmetries)~\cite{Nijholt2016,Nijholt2020,Makse2025} that also link to more general structural features such as graph partitions~\cite{Antoneli2006,schaub2016graph,Aguiar2018}. 
Computational algorithms can be used to identify possible cluster synchrony patterns~\cite{pecora2014cluster,kamei2013computation}; algorithms for pairwise networks remain applicable for higher-order structures via incidence matrices of hypergraphs~\cite{Aguiar2020}.
Stability properties of cluster synchrony are also restricted by structural features. 
For example, symmetry restricts the spectrum of the linearization, which has consequences for the stability and bifurcations of cluster synchrony as (partially) symmetric states~\cite{Golubitsky2002}.
To determine stability numerically, one may compute the irreducible representations of symmetry groups~\cite{pecora2014cluster} or the finest simultaneous block diagonalization of matrices in the variational equation~\cite{zhang2020symmetry}.
These techniques identify invariant subspaces of the dynamics to reduce the dimensionality of the stability problem and have been generalized to synchronization patterns on higher-order networks~\cite{zhang2021unified,salova2021cluster,salova2021analyzing}.

Further general collective dynamical phenomena of interest in higher-order network dynamics include heteroclinic cycles~\cite{Bick2023b}, clustering \cite{kovalenko2021contrarians}, and chimera states~\cite{li2023chimera,muolo2024phase,wang2024multi,muolo2024pinning,djeudjo2025chimera}.
For example, by treating chimera states as a special cluster synchronization pattern, simultaneous block diagonalization techniques have been used to characterize the stability of chimeras in the presence of nonpairwise interactions~\cite{zhang2021unified}.

\subsection{Phase reduction for periodic intrinsic dynamics}
\label{sec:phase_reduction}

If the dynamics of an isolated node in \cref{eq:coupled-oscillators-HOI} are periodic---each isolated node with state $x_i\in\R^d$ is an oscillator $\dot{x}_i = F_i(x_i)$ with dynamics on a stable limit cycle---
its synchronization dynamics can be captured by reducing the system to a phase description.
That is, the state of oscillator~$i$ is determined solely by a phase variable~$\theta_i\in \T = \R/2\pi\Z$ on the circle.
Probably the most famous phase description is the Kuramoto model~\cite{kuramoto1984chemical,Acebron2005} and its generalizations.
But also phase equations with nonpairwise interaction terms, such as $\sin(\theta_j+\theta_k-2\theta_i)$ involving the phases of oscillators~$i,j,k$, can be obtained through such \emph{phase reduction} from coupled nonlinear oscillators.

\begin{figure*}
\includegraphics[width=0.8\textwidth]{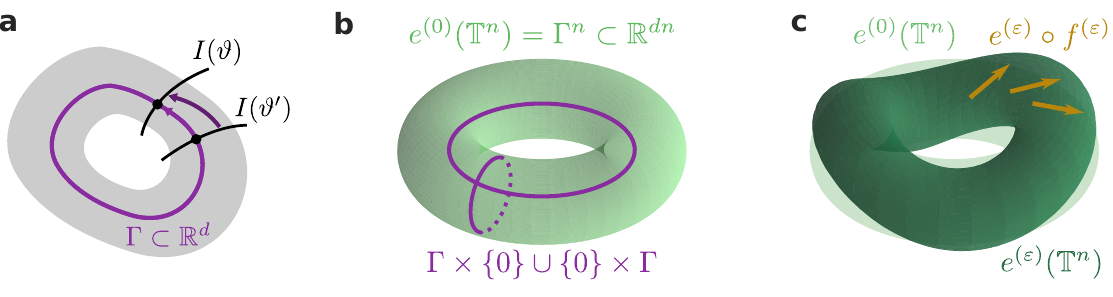}
\begin{minipage}{0.8\textwidth}
    \begin{justify}    
        The main intuition behind phase reduction is that the state on a (hyperbolic) limit cycle~$\Gamma\in\R^d$ can be described solely in terms the phase~$\vartheta\in\T$ of the oscillation~\cite{nakao2016phase,pietras2019network}; cf.~Panel~(a).
        In other words, there is a map $e_*^{(0)}:\T\to\R^d$ that assigns each phase the corresponding point on~$\Gamma$ and $\Gamma=e^{(0)}_*(\T)$.
    One typically defines~$\vartheta$ such that, in the absence of input, it evolves at uniform speed~$\omega$.        
        The notion of phase can be extended into the basin of attraction of the limit cycle; an isochron~$I(\vartheta)$ are the points of identical (asymptotic) phase.
        External forcing (e.g., through network interactions) affects the phase.
        The result are phase-reduced equations such as~\eqref{eq:OscPhaseRed}.

        For~$n$ oscillators (to keep notation simple assume they are identical) as an autonomous system---such as~\eqref{eq:OscPhys}---one can interpret a phase reduction as dynamics on an invariant torus; cf.~Panel~(b).
        Indeed, the uncoupled limit cycles ($\varepsilon=0$) form an (low-dimensional) invariant torus~$\Tb^{(0)} = \Gamma^n$ that attracts all nearby points.
        The map $e^{(0)}:\T^n\to\R^{nd}, (\theta_1, \dotsc, \theta_n)\mapsto (e^{(0)}_*(\theta_1), \dotsc, e^{(0)}_*(\theta_n))$ assigns each phase combination the corresponding point in phase space and $\Tb^{(0)} = e^{(0)}(\T^N)$.
        In other words, the dynamics reduce to~$\Tb^{(0)}$ and are described only by the phase variables; the ``amplitudes'' are determined by the phases.
        
        The phase description remains valid as the coupling strength is increased; cf.~Panel~(c).
        Fenichel's theorem~\cite{Fenichel1972} implies that the torus will persist up to a certain coupling strength~$\varepsilon_0$, that is, for $0<\varepsilon<\varepsilon_0$ there is a torus~$\Tb^{(\varepsilon)}$ close to~$\Tb^{(0)}$ which attracts all nearby points\footnote{
        As the coupling strength is increased invariant torus may eventually break down---but particular states, such as synchrony, may persist beyond the torus breakdown.}; this corresponds to a phase reduction.
        Computing the perturbed torus~$\Tb^{(\varepsilon)} = e^{(\varepsilon)}(\T^n)$ and the dynamics thereon given by a vector field~$f^{(\varepsilon)}:\T^n\to\R^{n}$ provides a way to compute a phase reduction~\cite{VonderGracht2023a}.        
        Specifically, expanding the phase dynamics~$f^{(\varepsilon)} = \omega+\varepsilon f^{(1)}+\dotsc$ and the embedding $e^{(\varepsilon)} = e^{(0)} + \varepsilon e^{(1)}+\dotsc$ one can compute~$e^{(\varepsilon)}, f^{(\varepsilon)}$ simultaneously order-by-order to the desired order.
        As a description for the dynamics, a phase reduction is not unique: 
        There is a trade-off between choosing coordinates on the torus (through~$e^{(\varepsilon)}$) and the phase dynamics~$f^{(\varepsilon)}$ so that they match the unreduced system~\eqref{eq:OscPhaseRed}.
        So when doing a phase reduction, one typically has to decide whether to preserve the meaning of phase of the uncoupled system (as in traditional approaches) or whether to reparameterize phases for the phase interactions to be as simple as possible (in normal form as in~\cite{VonderGracht2023a}).
        This approach can be extended to oscillators with time-delayed interactions~\cite{Bick2024}.

\end{justify}
\end{minipage}
    \caption{BOX PANEL: Phase descriptions of coupled oscillators.}
    \label{fig:MathBox}
\end{figure*}

If the coupling is sufficiently weak, then the collective dynamics of~$n$ coupled oscillators with high-dimensional state $x=(x_1, \dotsc, x_n)\in\R^{dn}$ 
can be reduced to the evolution of phases~$\theta = (\theta_1, \dotsc, \theta_n)\in \T^n$.
Written compactly, the dynamics of
\begin{align}
\label{eq:OscPhys}
\dot x_i &= F_i(x_i) + \eps G_i(x)
\intertext{are captured by the phases 
and their interactions}
\label{eq:OscPhaseRed}
\dot \theta_i &= \omega_i + \eps g^{(\eps)}_i(\theta),
\end{align}
see Box~\ref{fig:MathBox} for more mathematical details.
From a physical perspective, the idea is to parameterize the phase of each oscillator such that it evolves at constant speed~$\omega_i$ when uncoupled ($\eps=0$), extend the notion of phase into a neighborhood of the periodic orbit, and determine how the \emph{physical interactions~$G_i$} in~\cref{eq:OscPhys} affect the \emph{phase interactions~$g^{(\eps)}_i$} in~\cref{eq:OscPhaseRed}.
From a mathematical perspective, a phase reduction can be related to persistence of normally hyperbolic invariant manifolds~\cite{Fenichel1972}.
The phase interactions~$g^{(\eps)}_i$ may now contain nonpairwise interaction terms depending on the physical coupling~$G_i$.
Phase reductions are typically computed using an asymptotic expansion in the coupling strength~$\eps$.
While traditional phase reductions focus on the first-order expansion~\cite{nakao2016phase,pietras2019network}, the focus has shifted to compute second- and higher-order phase reductions~\cite{leon2019phase,mau2023phase,gengel2020high,bick2024higherorder,VonderGracht2023a}.

There are several ways how nonpairwise interactions can appear in the phase dynamics~\eqref{eq:OscPhaseRed} depending on the coupling of the nonlinear oscillators in \cref{eq:OscPhys}:
First, if the coupling in~$G_i$ of the nonlinear oscillators is nonpairwise, 
then one can expect the phase interactions to be nonpairwise at first order~\cite{ashwin2016Hopf,Len2025}.
Second, even when the coupling terms in~$G_i$ are pairwise, one may expect nonpairwise interactions to enter the phase dynamics at higher-order expansions~\cite{leon2019phase,bick2024higherorder}.
Intuitively, the emergent nonpairwise interactions capture indirect interactions between oscillators (e.g., that oscillator~$i$ receives input from oscillator~$k$ via~$j$ independent of any direct interaction between~$i$ and~$k$).

The functional form of the nonpairwise interactions depends on the coupling strength, the ``shape'' of the periodic orbit, and the physical interactions between the oscillators.
As an example, the phase interactions computed by Ashwin and Rodrigues~\cite{ashwin2016Hopf} for identical oscillators in \cref{eq:OscPhys} yield phase equations of the form%
\begin{equation}\label{eq:PhaseRed}
\begin{split}
\dot {\theta}_{k} &= \omega + \eps\sum_{j=1}^N  g_2(\theta_j-\theta_k)
\\& \qquad\qquad +\eps\sum_{j,l=1}^{N} g_3(\theta_j+\theta_{l}-2\theta_k)\\& \qquad\qquad
+\eps\sum_{j,l=1}^{N} g_4(2\theta_j-\theta_{l}-\theta_k)
\\& \qquad\qquad + \eps\sum_{j,l,m=1}^{N}  g_5(\theta_j+\theta_{l}-\theta_{m}-\theta_k)
\end{split}
\end{equation}
for $2\pi$-periodic coupling functions~$g_2,g_3,g_4,g_5$ (with one or two harmonics) that mediate pairwise, two types of triplet, and one type of quadruplet interactions.
Conversely, one may ask whether there are coupled nonlinear oscillators that have a given phase reduction.
For pairwise interactions, the goal of \emph{synchronization engineering}~\cite{Kori2008,Kiss2018} is to design oscillator coupling through feedback to achieve a target phase reduction.
A general theory for nonpairwise interactions is under development.

From the phase reduction perspective, when do the emergent nonpairwise interactions matter for the observed dynamics?
First, oscillators only interact in phase if they are resonant~\cite{Luck2011}, that is, their angular frequencies are in a rational relation; this applies for pairwise as well as nonpairwise interactions and relates to averaging~\cite{Swift1992,Sanders2007} and normal forms~\cite{VonderGracht2023a}.
Second, nonpairwise interactions that arise from higher-order phase reductions become more relevant as the coupling strength becomes larger and a first-order approximation breaks down~\cite{Ocampo-Espindola2024}.
Third, higher-order interactions play a role in determining the bifurcation behavior of a system: At bifurcation points, higher-order terms can determine the type of bifurcation~\cite{kuehn2021universal}.
Fourth, nonpairwise interaction terms matter when the dynamics of an approximation with pairwise coupling is degenerate.
For example, the Kuramoto-type interactions as a low-order approximation for a globally-coupled network of identical oscillators is degenerate~\cite{watanabe1994constants}, whereas a better approximation~\eqref{eq:PhaseRed} with nonpairwise coupling can reveal the possibility of chaotic dynamics~\cite{bick2016chaos}.

Phase reduction provides a direct link between coupled nonlinear oscillators and phase oscillator systems with nonpairwise interactions and dynamical phenomena that arise in both systems, including synchrony~\cite{bick2024higherorder} and chimeras~\cite{Mau2024}.
This provides an opportunity to link the perspective for phase oscillators with higher-order interactions as models \textit{per se}---such as insights from the underlying hypergraphs or simplicial complexes---with the emergent collective dynamics of coupled nonlinear oscillators.

\section{Reduction and Reconstruction of Higher-Order Networks from Data}

Given a higher-order structure such as a hypergraph or simplicial complex, one can define a dynamical system on it as discussed above.
But given dynamics, one can ask about what (higher-order) network coupling structure provides a good representation for the dynamics and data measured from it~\cite{Torres2021}.
For example, should it be a simplicial complex, a hypergraph, or a more general object such as a directed hypergraph~\cite{gallo2022synchronization,Bick2023b,zhang2023higher}?
This question has natural implications when there are different combinatorial objects for the same dynamics~\cite{Aguiar2020}.
Phase reduction (\cref{sec:phase_reduction}) provides an explicit example, where the order of the ``physical'' network interactions may be different from the ``effective'' phase interactions (pairwise vs. higher-order phase interactions).
This also has implications for network reconstruction~\cite{de2010advantages}; in the context of phase reduction, whether one reconstructs physical interactions or effective phase interactions.
Finally, what the best way to encode higher-order network structure is also depends on the question. 
For example, classical techniques applied to the incidence graph as a representation for a hypergraph can give insights on possible synchrony patterns~\cite{Aguiar2020}.

\subsection{How high an order is high enough?}
\label{sec:HighEnough}

\begin{figure}
    \centering
    \includegraphics[width=0.8\linewidth]{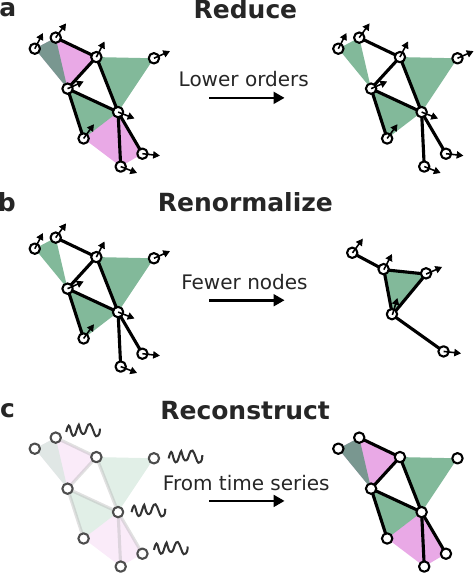}
    \caption{One can decrease the complexity of a higher-order network by either (a) reducing the maximum order of interactions, or (b) coarse grain it into fewer nodes. Both reduction and renormalization utilize information from observing the dynamics unfolding on the nodes. Complementarily, when the coupling structure is unknown to begin with, there are methods to reconstruct it from observed time series (c).}
    \label{fig:reduction}
\end{figure}

Including higher-order interactions yields additional combinatorial complexity: The computational overhead of models and algorithms increase exponentially with the maximum interaction order being considered.
So what is the minimal order necessary to represent observed dynamics (\cref{fig:reduction}a)?
For cluster synchronization, there are explicit estimates for the minimal order necessary~\cite{VonderGracht2023}.
More generally, one can break down interactions depending on the form and nonlinearity of the interaction functions~\cite{aguiar2011dynamics} and identify systems that are dynamically equivalent.
Along similar lines, one can determine conditions under which some dynamical processes on hypergraphs can be exactly rewritten as dynamics on a new hypergraph with a lower maximum order~\cite{neuhauser2024learning} by contrasting the \textit{topological order} $d_{\rm topo}$ of a hypergraph---determined by the size of the largest hyperedge---with the \textit{dynamical order} $d_{\rm dyn}$ determined by the coupling functions of the dynamics.

For example, take a hypergraph with dyadic and triadic interactions, so that $d_{\rm topo}=3$. Now, if the triadic functions are $g_3(\theta_i, \theta_j, \theta_k) = \sin(\theta_j + \theta_k - 2 \theta_i)$, they cannot be linearly decomposed into functions of fewer phases, and hence $d_{\rm dyn}=3$. 
However, if $g_3(\theta_i, \theta_j, \theta_k) = \sin(\theta_j - \theta_i) + \sin(\theta_k - \theta_i)$, then $d_{\rm dyn}=2$, because it is the linear combination of two dyadic coupling functions. Finally, these two orders can be combined into an \textit{effective order} $d_{\rm eff}$ bounded by $d_{\rm eff} \le \min ( d_{\rm topo}, d_{\rm dyn})$, where the equality holds most of the time. Note that this framework only works for coupling functions~$f_d$ that are invariant under any permutation of their last $d-1$ arguments. In higher-order Kuramoto dynamics, this would include $\sin(\theta_j + \theta_k - 2 \theta_i)$ but not $\sin(2 \theta_j - \theta_k - \theta_i)$.

The question can also be approached from the perspective of model selection~\cite{lucas2026reducibility,kirkley2025structural}. Given a hypergraph with maximum interaction order $d_{\rm max}$, when is a reduced hypergraph with hyperedges up to $d<d_{\rm max}$ an optimal model of the original hypergraph? 
To do this, beyond purely topological arguments~\cite{kirkley2025structural}, one can compare the density matrices of reduced hypergraphs at each order with that of the original hypergraph~\cite{lucas2026reducibility}, which are representations of higher-order diffusion processes on the hypergraphs, at a chosen diffusion time.
The quality of a given reduced hypergraph can then be assessed with a cost function: better reduced hypergraphs minimize the cost function by maximizing model accuracy while minimizing model complexity, in the spirit of the minimum message length framework.
The method determines an optimal order $d_{\rm opt} < d_{\rm max}$ at which the hypergraph can be truncated, discarding higher orders, effectively compressing the original hypergraph. The results indicate that, while some systems are fully reducible to only dyadic interactions, others cannot be reduced at all.

\subsection{Renormalizing to smaller hypergraphs}

In the above studies, the goal was to reduce the system by discarding large hyperedges while preserving the dynamics of the system. 
Another way of reducing the complexity of a system is to reduce the number of nodes by merging similar nodes, following the classical ideas of the Renormalization Group (RG)~\cite{kadanoff1966scaling} (\cref{fig:reduction}b). 
Importantly, in scale-invariant systems, this zooming-in should not affect the dynamics of the system. 
Using metric space embedding, multiple approaches have been proposed to extend the RG framework to (pairwise) complex networks~\cite{serrano2008selfsimilarity,garcia-perez2018multiscale}. 
It is only recently that a study proposed to use diffusion to naturally capture the topological---rather than geometric---aspect of the problem~\cite{villegas2023laplacian}. 
Nurisso \textit{et al.}\ then generalized the approach to higher-order networks by defining a cross-order Laplacian~$\mathrm{L}^{\times}_{(d_1, d_2)}$ that captures diffusion between interactions of order $d_1$ and $d_2$~\cite{nurisso2025higherorder} (see \cref{sec:beyond} for hyperedge-based approaches). 
Similar to~\cite{lucas2026reducibility}, the authors use the density matrix formalism, but this time to merge nodes into super-nodes and to define a higher-order \textit{scale-invariance parameter} to quantify scale-invariance. 
Results showed that empirical systems from various domains displayed different higher-order scale-invariance profiles. 
Utilizing an idea distinct from the RG-based approaches, Thibeault \textit{et al.}~\cite{thibeault2024lowrank} recently showed that many complex systems can be reduced to a low-rank matrix by keeping only the most relevant singular values from the original adjacency matrix of the system. Importantly, higher-order interactions can emerge when applying this reduction to pairwise networks.

\subsection{Inferring higher-order structures from dynamics}

Given dynamics, the inverse problem of inferring the organization of higher-order networks from time series is equally important (\cref{fig:reduction}c).
This is especially relevant given that direct measurements of higher-order interactions in many complex systems are challenging with current experimental techniques.
For networks with pairwise connections, network inference problems have a long history~\cite{de2010advantages}.
Inferring higher-order interactions from data has been less studied but is a rapidly developing field~\cite{battiston2021physics,wang2022full,wang2022supervised,wegner2024nonparametric,tabar2024revealing,li2024higher,baptista2024mining,delabays2024hypergraph}.
Here, we will focus mainly on dynamical systems approaches for hypergraph reconstruction~\cite{kralemann2011reconstructing,kralemann2013detecting,kralemann2014reconstructing}.
However, it is worth noting that other approaches such as information-theoretic techniques have also been shown to be effective and can reveal insights complementary to the dynamical systems approaches~\cite{young2021hypergraph,lizotte2022hypergraph,contisciani2022inference,varley2023partial,santoro2023higher,santoro2024higher}.

A common setup for hypergraph reconstruction is based on \cref{eq:coupled-oscillators-HOI}. We assume that the adjacency tensors~$A_{ij}$ and $B_{ijk}$ are unknown and would like to infer them from observed trajectories of~$x$.
If the intrinsic dynamics~$F_i$ and the coupling functions $G^{(2)}$ and $G^{(3)}$ are known, Ref.~\cite{malizia2024reconstructing} showed that one can map the hypergraph reconstruction problem to linear matrix equations, which can be solved using optimization techniques such as ordinary least squares, Signal Lasso, or non-negative least squares.
Because the number of hyperedges that need to be considered grows rapidly with the system size and interaction order, it can become computationally challenging to reconstruct large hypergraphs.
Ref.~\cite{zang2024stepwise} further improved the computational efficiency of the method by focusing on important special cases such as systems with weak higher-order interactions. 

When the underlying dynamics and/or coupling functions are unknown, we need model-free inference methods~\cite{casadiego2017model}, which can be especially effective for complex systems whose precise dynamics are difficult to model.
One recent idea is to perform the Taylor expansion of \cref{eq:coupled-oscillators-HOI} around an arbitrary base point~\cite{delabays2024hypergraph}. 
This provides the theoretical basis for applying sparse regression techniques~\cite{brunton2016discovering} with monomial feature libraries to observed trajectories and look for terms such as $x_i x_j x_k$ in the identified equation, which indicates the existence of triadic interactions among the nodes~$i$, $j$, and~$k$. 

Phase reduction also provides a powerful tool for reconstructing hypergraphs.
For example, Refs.~\cite{kralemann2011reconstructing,kralemann2013detecting,kralemann2014reconstructing} used phase reduction and spectral decomposition to infer the \textit{effective} connectivity between the phase-reduced oscillators. 
The method involves numerical estimation of their phase, derivative, and coupling functions approximated by Fourier series, the coefficients of which are associated with the weights of the reconstructed directed hyperedges.
More generally, if there are multiple representations of the same dynamics (e.g., through phase reductions), there is a question whether network reconstruction captures physical or (higher-order) effective interactions~\cite{Nijholt2022}.

The techniques above all require estimating the derivatives from data, which can be sensitive to noise and imposes constraints on how far apart adjacent data points can be.
A promising research direction is to incorporate advances in derivative-free methods~\cite{chen2018neural} to make hypergraph reconstruction more robust to noise and applicable to sparsely sampled data.
Computational cost is another issue that can be further improved.
Despite solid progress in making the inference more efficient, currently it is still challenging to reconstruct interactions beyond the third order for general hypergraphs with over a few hundred nodes.
New ideas are needed to scale up the inference to thousands of nodes and beyond. 

Finally, future works have the opportunity to develop and apply hypergraph inference methods to real-world data. 
Taking this challenge head-on can have a significant impact on fields such as ecology and neuroscience~\cite{varley2023partial,santoro2023higher,santoro2024higher}.
For example, if we think of the brain as an interconnected dynamical system, are networks an adequate model to describe the couplings between brain regions, or are nonpairwise interactions needed to capture the observed brain dynamics?
By applying the model-free hypergraph inference method from Ref.~\cite{delabays2024hypergraph} to resting-state EEG data, it was found that nonpairwise interactions can contribute significantly to macroscopic brain dynamics.

\section{Beyond node dynamics: Dynamical Higher-Order Networks}
\label{sec:beyond}

In the previous sections we have made the implicit assumption that the dynamics of the higher-order networks take place exclusively on the nodes. 
Alternatively, one can associate dynamical variables with edges and hyperedges to represent fluxes in transport networks such as the ocean~\cite{schaub2020random}, 
opinions of groups of individuals~\cite{sampson2025oscillatory},
or the brain's dynamic functional connectivity~\cite{santoro2023higher}. 
Such dynamics have mostly been considered in the context of simplicial complexes, where one can exploit the rich theory from discrete geometry and topology; see Refs.~\cite{nurisso2023unified,millan2025topology} for recent reviews on such types of edge-based dynamics.

\begin{figure}[tb]
    \centering
    \includegraphics[width=.96\linewidth]{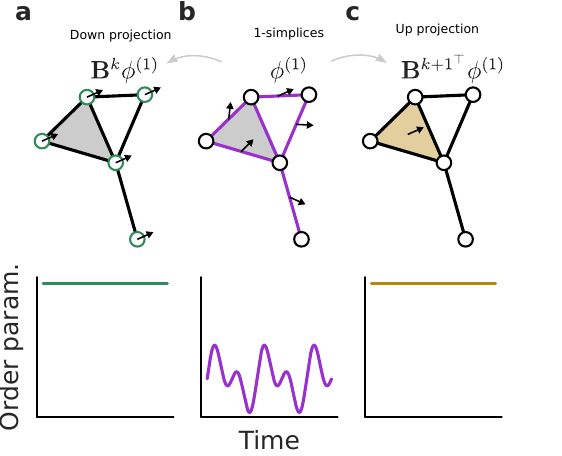}
    \caption{Simplicial Kuramoto models: dynamics of phases associated to $k$-simplices. For example, the dynamics of (b) 1-simplices, projected (a) down to 0-simplices and (c) up to 2-simplices. The dynamics of the order parameter depends on the order of interaction.}
    \label{fig:beyond_nodes}
\end{figure}

The simplicial Kuramoto model, also known as the topological Kuramoto model~\cite{millan2020explosive}, describes the synchronization dynamics of oscillator phases~$\phi^{(k)}\in\T^{d(k)}$ placed on the $k$-simplices of a simplicial complex. 
The model is illustrated in \cref{fig:beyond_nodes} and elegantly formulated with boundary ($\mathbf{B}^{k}$) and coboundary ($\mathbf{B}^{k^\top}$) operators describing the adjacency relations between simplices as
\begin{equation*}
\dot{ {\phi}}^{(k)} = \omega  
- \sigma^\uparrow\mathbf{B}^{k+1} \sin(\mathbf{B}^{k+1^\top}\phi^{(k)}) 
-\sigma^\downarrow \mathbf{B}^{k^\top} \sin(\mathbf{B}^{k}\phi^{(k)})\; .     
\end{equation*}
For $k=0$, the last term vanishes and the model reduces to the standard Kuramoto model, but for $k>0$, two different types of interactions emerge: one from below ($\downarrow$) involving adjacent lower-dimensional faces ($k-1$), and one from above ($\uparrow$) involving higher-dimensional faces ($k+1$).
Interactions from above involve $k+2$ oscillators, and thus for $k>0$ are genuinely higher-order: 
They cannot be reduced to a combination of pairwise interactions.
The functional form of the interactions from below depends on the number and direction of $k$-simplices adjacent to each $(k-1)$-subface, and include self-interactions from free subfaces, genuinely higher-order interactions when more than two simplices are adjacent to a subface, and pairwise interactions  
\cite{nurisso2023unified}.

The topological Kuramoto model leads to a continuous synchronization transition for any $k$, as was initially shown through computational analyses~\cite{millan2020explosive}.
Necessary and sufficient conditions for the existence and stability of phase-locked states have recently been attained by writing the model as a gradient flow~\cite{nurisso2023unified}.  

The topological Kuramoto dynamics exhibit fundamental differences from the node-based  model, revealed through the discrete Hodge Laplacian $\mathbf{L}^{k} = \textbf{L}_\downarrow^k + \textbf{L}_\uparrow^k = \textbf{B}^{k-1^\top}\textbf{B}^k + \textbf{B}^{k+1} \textbf{B}^{k^\top}$ describing the linearized dynamics. 
By the simplicial Hodge decomposition theorem, one can see that divergence-free and curl-free components evolve independently via $\textbf{L}_\uparrow$ and $\textbf{L}_\downarrow$, respectively, 
while harmonic modes remain stationary~\cite{millan2020explosive}. 
Phase coupling is thus possible only when the harmonic part of the natural frequencies vanishes and 
the harmonic eigenvectors localize along topological holes in the complex, making these structural features the key drivers of synchronization~\cite{nurisso2023unified}: 
Without holes the dynamics freezes, but when present, synchronization evolves along these localized modes.
Consequently, global synchronization requires specific topological conditions—such as a single delocalized hole as found in torus tessellations~\cite{carletti2023global}—that are fundamentally different from the connectivity requirements in standard node-based models.
Alternatively, one can re-define phase-locking as a state such that the curl-free and divergence-free components freeze, which respects the model symmetries~\cite{millan2020explosive, arnaudon2021connecting, nurisso2023unified}.

A central question is whether a link exists between topological and node-based Kuramoto dynamics. 
Ref.~\cite{nurisso2023unified} shows that topological Kuramoto dynamics of order $k$ is equivalent to higher-order node Kuramoto dynamics on an effective hypergraph where nodes correspond to the original $k$-simplices and hyperedges encode topological coupling. Thus, the simplicial Kuramoto model can be seen as a particular kind of node-based Kuramoto dynamics where the coupling functions depend on the orientations of the original simplices and do not vanish when all phases are equal.
For simplicial manifolds, this reduces to the standard pairwise node Kuramoto model. 
That being said, the edge-based formulation often helps analytical treatment through topology and discrete geometry tools when the system is originally a simplicial Kuramoto.
In addition, it enables the study of the topological Kuramoto model as a general system of higher-order coupled oscillators operating in a near-resonant regime~\cite{nurisso2023unified}.

Many variations of these topological Kuramoto models are possible.
On the one hand, there are other variants such as the ``{adaptive}'' or ``{explosive}'' topological Kuramoto model, which exhibit explosive discontinuous transitions~\cite{millan2020explosive,ghorbanchian2021higher}, while \emph{Hodge-coupled} formulations give rise to bistable regimes of phase and anti-phase synchronization~\cite{nurisso2023unified}.
Frustration may also be introduced via the \emph{simplicial Sakaguchi-Kuramoto model}~\cite{arnaudon2021connecting} by adding phase lags.
On the other hand, the dynamics across different dimensions of the simplicial complex may also be coupled using the topological Dirac operator~$\textbf{D}$, which has a tridiagonal structure built from boundary operators that enables cross-talk between signals of different dimensions.
Given that $\textbf{D}^2 = \text{diag}(\textbf{L}^0, \dotsc, \textbf{L}^k)$, the Dirac operator is often viewed as the square root of the (block-diagonal) Hodge Laplacian~\cite{bianconi2021topological, muolo2024three}.
This operator naturally couples node and edge signals locally, resulting in discontinuous synchronization transitions with the emergence of spontaneous rhythms~\cite{calmon2023local}.
The Dirac framework can be extended to couple orders adaptively—including coupling adjacent orders or global coupling across all orders—in an \emph{explosive Dirac-Kuramoto} model~\cite{ghorbanchian2021higher, nurisso2023unified}.

Apart from synchronization phenomena in topological Kuramoto networks, dynamical higher-order networks also arise if the higher-order interactions are subject to adaptation; see~\cite{Berner2023} for a recent survey on adaptive networks.
On the one hand, adaptive higher-order interactions and how they affect synchronization transitions have been considered as generalizations of Kuramoto oscillator networks~\cite{Kachhvah2022,Sharma2024,Rajwani2023}.
On the other hand, adaptivity~\cite{li2025bounded} of higher order interactions can also shape the dynamics of voter models and opinion formation~\cite{Horstmeyer2020,Golovin2024} or multiplayer game dynamics~\cite{Schlager2022}.

\section{Outlook and Perspectives}

In this review, we summarized key results and emerging directions in the study of collective dynamics on higher-order networks. 
Results from the last few years indicate that a variety of novel and rich behaviors emerge when higher-order interactions are taken into account, such as the promotion of explosive phase transitions or the emergence of new stable states, which cannot be achieved with only dyadic couplings.

The effects of higher-order interactions, such as explosive transitions or multistability, have been observed in a variety of dynamical processes including contagion and rumor spreading dynamics. When relevant, we discussed the generality of these phenomena across dynamical processes. For more focused discussion on contagion and simplicial dynamics, see the following recent reviews \cite{ferraz2024contagion,millan2025topology}.

While a few analytical schemes are in principle designed to tackle general higher-order networks with interactions of any order, in practice most research so far has focused on the investigation of novel phenomena when only two-body and three-body interactions are considered. 
Recent studies on higher-order Ising model~\cite{robiglio2024higher,son2024phase} and complex contagion~\cite{kiss2023insights} show that further novel behaviors might emerge when interactions of even higher order (e.g., from four-body onward) are considered. 
An interesting future direction 
is to investigate the potential for new collective phenomena emerging beyond three-body interactions.

Another promising direction is to study the effect of different nonpairwise coupling functions.
Interaction functions obtained through phase reduction (\cref{sec:phase_reduction}), provide natural classes of interaction functions that link to the dynamics of nonlinear oscillators.
But from the perspective of network dynamics on higher-order networks as models \emph{per se}, there are a few ``standard'' coupling functions that are regularly considered for Kuramoto oscillators, but there is little consensus on what nonpairwise coupling functions to use for general oscillator dynamics.
Even for the symmetric and asymmetric coupling functions in \cref{eq:kuramoto}, researchers often pick one or the other based on technical convenience, and their effects on collective dynamics are not yet fully understood. 
In particular, it remains an open problem to understand which higher-order coupling functions are better able to support synchronization, 
what kind of multistability arise from different coupling functions,
and if new collective phenomena may emerge from the mixture of multiple coupling functions.

Moving forward, it is also important to improve our understanding of when techniques for pairwise networks can be adapted to higher-order networks and when they fail.
In some cases, straightforward modifications of the network approach are possible. 
This includes multi-order Laplacians for linear stability analyses of synchronization on Kuramoto oscillators and finding admissible synchronization patterns via a hypergraph incidence matrix rather than an adjacency tensor.
In other cases, fundamentally new approaches need to be developed.
For example, the Ott--Antonsen ansatz works for the standard Kuramoto model, but fails when the symmetric three-body coupling function in \cref{eq:kuramoto} is introduced.

While higher-order interactions can yield new collective dynamical phenomena, we pay a price with the additional combinatorial complexity.
Thus, rather than considering dynamics of any order, a change of perspective can provide a way forward.
On the one hand, it is critical to understand when higher-order interactions are necessary.
For example, for a given dynamical transition, what is the order required to see this phenomenon (cf.~\cref{sec:HighEnough})?
This allows to restrict to models of a certain maximal order.
On the other hand, one needs to identify equivalences between different combinatorial models for network interactions---possibly of different order (see~\cref{sec:phase_reduction}). 
This will allow us to use the right language (e.g., graphs vs hypergraphs) and coordinates (e.g., phase-reduced dynamics vs unreduced dynamics) to analyze the dynamical phenomenon at hand.

Finally, informed by the availability of increasingly rich empirical data, higher-order models of collective dynamics have great potential to advance our understanding in fields such as ecology and neuroscience.
In ecology, dynamical systems techniques can give insights on how higher-order interactions shape resilience and diversity of ecosytems~\cite{Levine2017,Gibbs2022}.
In neuroscience, studies of the multi-way correlations between brain signals~\cite{neri2025taxonomy,geli2025nonequilibrium,santoro2024higher,rosas2022disentangling,majhi2025patterns,santoronodes,santoro2025beyond} 
have led to new insights about the inner workings of the brain.

\begin{acknowledgments}
F.B. acknowledges support from the Austrian Science Fund (FWF) through project 10.55776/PAT1052824 and project 10.55776/PAT1652425. 
C.B. acknowledges support by the project ``BeyondTheEdge: Higher-Order Networks and Dynamics'' (European Union, REA Grant Agreement No.~101120085).
M.L. is a Postdoctoral Researcher of the Fonds de la Recherche Scientifique-FNRS.
Y.Z. acknowledges support from the Santa Fe Institute Omidyar Fellowship and the National Science Foundation (NSF DMS 2436231).
\end{acknowledgments}

\section*{Data availability}

To facilitate the study of dynamics on higher-order networks, with this review, we release \texttt{hypersync}, an open-source Python package for the simulation, analysis, and visualization of oscillators with higher-order interactions, available at \url{https://github.com/maximelucas/hypersync}.

\bibliography{bibli}

\end{document}